\documentclass[conference]{IEEEtran}

\ifCLASSINFOpdf
\else
\fi
\usepackage{amssymb}
\usepackage{dsfont}
\usepackage{multirow}
\usepackage{amsmath}    
\usepackage{graphicx}   
\usepackage{verbatim}   
\usepackage{color}      
\usepackage{subfigure}  
\usepackage{algorithmic}
\usepackage{algorithm}
\usepackage{float}

\begin{document}
%
\title{On the Privacy of Two Tag Ownership Transfer Protocols for RFIDs}

\author{\IEEEauthorblockN{Mohammad Reza Sohizadeh Abyaneh}
\IEEEauthorblockA{Selmer Center\\
University of Bergen\\
reza.sohizadeh@ii.uib.no}}

\maketitle

\begin{abstract}
In this paper, the privacy of two recent RFID tag ownership transfer protocols are investigated against the tag owners as adversaries.\\
The first protocol called ROTIV is a scheme which provides a privacy-preserving ownership transfer by using an HMAC-based authentication with public key encryption. However, our passive attack on this protocol shows that any legitimate owner which has been the owner of a specific tag is able to trace it either in the past or in the future. Tracing the tag is also possible via an active attack for any adversary who is able to tamper the tag and extract its information.\\
The second protocol called, \emph{Chen et al.}'s protocol, is an ownership transfer protocol for passive RFID tags which conforms EPC Class1 Generation2 standard. Our attack on this protocol shows that the previous owners of a particular tag are able to trace it in future. Furthermore, they are able even to obtain the tag's secret information at any time in the future which makes them capable of impersonating the tag.
 \end{abstract}

\IEEEpeerreviewmaketitle

\section{Introduction}
Radio frequency identification(RFID) is currently considered as the next generation technology that mainly used to identify massive objects in an automated way and will substitute traditional optical barcode system in near future. The RFID advantages such as reducing supply chain inefficiencies and improving inventory flow leaves no doubt that the dominant deployment of barcodes nowadays in supply chain will be promptly taken over by RFID tags. But it has its own drawbacks too.\\
As products flow through a supply chain, their ownership is transferred from one partner to the next. This transfer of ownership extends to the RFID tags
attached to these products. Thus all information associated with the tag will need to be passed from the current to the new owner. However, at the moment of tag ownership transfer, both the current and new owners have the information necessary to authenticate a tag, and this fact may cause an infringement of tag \emph{owner privacy} \cite{Albert}.\\
To handle this problem, tag ownership transfer protocols are proposed to transfer the ownership of a tag from one owner to another securely. The proposed schemes for ownership transfer protocols are divided into two groups. Some schemes exploit a trusted third party(TTP) which acts as a secure channel to transfer some information between the entities. One of the first solution of this kind was proposed by Saito \emph{et al}.\cite{Saito}. However, the security of their scheme is only based on the short read range of the backward channel (tag to reader communication) by assuming that it is hard for adversaries to eavesdrop on this channel. Another scheme with TTP is proposed by Molnar \emph{et al.} \cite{Molnar}. They exploit the TTP to manage tag keys by a tree structure. But in this protocol one key is shared by several tags which makes this protocol vulnerable. The privacy of the whole system decreases quickly when more tags are compromised \cite{Avoine2}.\\
There also exist some \emph{decentralized} proposals without a using TTP. Most of these schemes have two following assumptions: there is a secure channel between the current and new owner to pass the tag's information securely. They also assume that the new owner and the tag will be able to execute an authentication session in an isolated environment without presence of the current owner after the ownership transfer is completed in order to update some secret parameters.\\
For instance, Soppera and Burbridge \cite{Soppera} adopt the scheme of Molnar \emph{et al}. by replacing the TTP with some distributed local devices called RFID \emph{acceptor} tag. In \cite{Yoon}, the authors have also proposed a decentralized protocol relying on the assumption that owners are able to change the tag key in an isolated environment. However, this protocol has security vulnerabilities well described in \cite{Kapoor}. Song \emph{et al.} \cite{Song} proposed a scheme with introduction of a new property called \emph{authorization recovery} which facilitates the ownership transfer of a tag to its previous owner. But Pedro \emph{et al.} \cite{Peris} showed that their schemes has some vulnerabilities as well.\\
Recently, two other tag ownership transfer protocols have been proposed. The first scheme is called an RFID ownership transfer with issuer verification (ROTIV) \cite{Elkhiyaoui} which provides a constant-time, privacy-preserving tag ownership transfer. The ROTIV's main idea is to combine an HMAC-based authentication with public key encryption. The second scheme which is proposed by Chen \emph{et al.} \cite{Chen3}, proposes an RFID ownership transfer systems which conforms the requirements of EPCglobal Class-1 Generation-2 Standard.\\
\textbf{Our Contribution. }In this paper, we investigate the privacy of two aforementioned ownership transfer protocols. The investigation includes some attacks to violate the forward and backward privacy as well as previous and new owner privacy properties of the schemes.\\
\textbf{Outline. }The remainder of this paper is organized as follows. Section \ref{prim} describes the privacy issues and properties required for tag ownership transfer protocols as well as system and adversary modelings. In Sections \ref{rotiv2} and \ref{chen2} the description of the the ROTIV and Chen \emph{et al.} protocols and our attack on them are presented respectively, and finally, Section \ref{conclusions} concludes the paper.
\section{Preliminaries}\label{prim}
To lend clarity to our discussions in the subsequent sections, in this section, we outline the models and properties used in ownership transfer protocol.
\subsection{System Model}
In ownership transfer protocols, there are mainly three active entities involved: \emph{current owner}, \emph{tag} and \emph{new owner}. The owners in an ownership transfer protocols are some readers in practice which take the role of ownership in these kinds of protocols. The ownership transfer protocols typically provide a solution to transfer the tag's information from the current owner to the new owner.\\
Most of the ownership transfer protocols consist of two phases, an \emph{authentication phase} and a \emph{ownership transfer phase}. By the former phase, the tag and two owners are mutually authenticated and the latter phase assures all three entities that the ownership of the tag is transferred in a proper and privacy-preserving way.
\subsection{Privacy Properties}\label{privacy}
Generic privacy properties and how to formalize them for RFID systems have been extensively explored in the literature \cite{Avoine,Vaudenay,Juels,Perneel}. The two generic privacy property we address in this paper are:
\begin{itemize}
  \item \emph{Backward Privacy}: an adversary should not be able to to trace past transactions between an owner and a tag, even if it compromises/tamper the
tag.
  \item \emph{Forward Privacy}: an adversary should not be able to to trace future transactions between an owner and a tag, even if it compromises/tamper the
tag.
\end{itemize}
On the other hand, in tag ownership transfer protocols changes of tag owner could occur frequently and at the moment of tag ownership transfer, both the current and new owners have the information necessary to authenticate a tag, and this fact may cause an infringement of tag \emph{owner privacy}. Therefore, there are two extra privacy issues dedicated for ownership transfer protocols in the literature \cite{Osaka,{Song2}}:
\begin{itemize}
\item \emph{New owner privacy}: Once ownership of a tag has been transferred to a new owner, only the new owner should be able to identify and control the tag. The previous owner of the tag should no longer be able to identify or \emph{trace} the tag.
\item \emph{Current/previous owner privacy}: When ownership of a tag has been transferred to a new owner, the new owner of a tag should not be able to \emph{trace} past interactions between the tag and its previous owner.
\end{itemize}
\subsection{Adversary Model}
 In \cite {Juels}, Juels and Weis give a formal model of the privacy in RFID systems. In this model, tags ($\mathcal{T}$) and readers/owners ($\mathcal{R}$) interact in protocol sessions. During this interaction there is also an adversary entity $\mathcal{A}$ which passively or actively interacts with them. The adversary may have access to an oracle which can be queried by the following queries:
\begin{itemize}
  \item \texttt{Execute}($\mathcal{T},\mathcal{R}, i$): This query is responded by the information of $\mathcal{T}$ and $\mathcal{R}$ interactions in an honest protocol session at time instance $i$.
  \item \texttt{Send}($\mathcal{P}_1,\mathcal{P}_2, i,m$): This query models active attacks by allowing the adversary $\mathcal{A}$ to impersonate some entity, a tag or a reader, $\mathcal{P}_1$ in some protocol session $i$ and send a message $m$ of its choice to an instance of some other entity $\mathcal{P}_2$.
\item \texttt{Corrupt}($\mathcal{T}$): This query allows the adversary $\mathcal{A}$ to tamper the tag to learn the stored secret information of the tag $\mathcal{T}$
  \item \texttt{Test}($i,\mathcal{T}_0,\mathcal{T}_1$): This query is responded by a random bit $b\in\{0,1\}$ and the interaction information of the tag $\mathcal{T}_0$ and $\mathcal{T}_1$ with the reader/owner at $i^{th}$ time instance.
\end{itemize}
\subsection{Attack Scenario}\label{scenario}
In \cite {Juels}, the adversary $\mathcal{A}$ aims at tracing a specific target tag $T$. To do so, she,
\begin{itemize}
  \item absorbs the information she requires about the target tag $T$ by the means of queries previously described.
  \item choose two test tags $T_0$ and $T_1$ where one of them is $T$, and asks the oracle for the challenge by \texttt{Test} query. The response will be the interactions between the $T_0$ and $T_1$ tags with the reader $R$ at a specific time instance.
\end{itemize}
The adversary succeeds to violate the privacy of the tag by tracing it, if she is able to distinguish the tag $T$ between the two tested tags by outputting 0 or 1.
 \subsection{Notations}
Here, we explain the notations used hereafter.
\begin{itemize}
\item $E_k(.)$: Symetric/asymetric encryption function operation with the key $k$.
\item $pk_X,sk_X$: Public and private key of entity $X$ respectively.
\item $h_k(.)$: Keyed hash function with key $k$.
\item $h(.)$: Hash functions.
\item $PRNG(.)$: Pseudo random number generator.
\item $T,O_n,O_{n+1}$: Tag, current owner and new owner.
\item $ID_X$: The identification (ID) of entity $X$.
 \item $N_X$: Random numbers generated by entity $X$.
 \item $m_i$: dynamic value $m$ at time instance $i$.
\end{itemize}
\section{ROTIV Protocol}\label{rotiv2}
ROTIV is a decentralized scheme which does not require a trusted third party to perform tag ownership transfer. This protocol provides issuer verification that allows prospective owners to check the identity of the entity which has issued the tag. The authors have claimed that their scheme ensures both forward and backward privacy and it also preserves current and new owner privacy.\\
There are four entities involved in the protocol, a tag $T$, current owner $O_n$, new owner $O_{n+1}$ and issuer $I$ which initializes the tag and owners.\\
In ROTIV, the $T$ stores a symmetric key $k$, a \emph{state parameter} $s$, where $k$ is a key shared between the tag and its owner and $s$ is an Elgamal encryption of $T$'s identification information.
\subsection{Preliminaries}
\textbf{Bilinear pairing}\\
Let $\mathbb{G}_1$, $\mathbb{G}_2$ and $\mathbb{G}_T$ be groups, such that $\mathbb{G}_1$ and $\mathbb{G}_2$ have the same prime order $q$. Pairing $e : \mathbb{G}_1 \times \mathbb{G}_2 \rightarrow \mathbb{G}_T$ is a bilinear pairing if has the following properties:
\begin{enumerate}
                   \item  \emph{bilinear}: $\forall a, b  \in  \mathds{Z}_q$ , $g_1 \in \mathbb{G}_1$ and $g_2 \in \mathbb{G}_2$ , $e(g^a_1,g^b_2) = e(g_1, g_2)^{ab}$.
                   \item \emph{computable}: there is an efficient algorithm to compute $e(g_1, g_2)$ for any $(g_1, g_2) \in \mathbb{G}_1 \times \mathbb{G}_2$;
                   \item \emph{non-degenerate}: if $g_1$ is a generator of $\mathbb{G}_1$ and $g_2$ is a
generator of $\mathbb{G}_2$, then $e(g_1, g_2)$ is a generator $\mathbb{G}_T$.
                 \end{enumerate}

\subsection{Description}
\begin{figure}[t!]
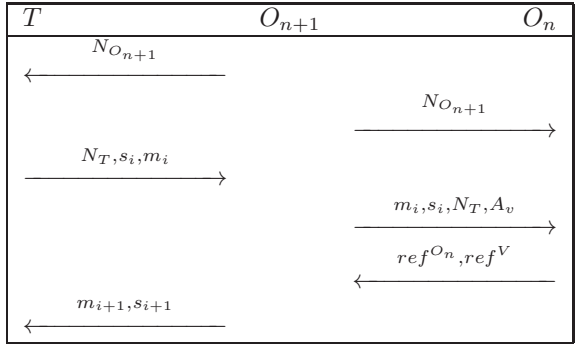

\begin{center}
\begin{tabular}{|l l r|}
\hline
\textbf{$T$}  & \textbf{$O_{n+1}$}   & \textbf{$O_{n}$} \\
\hline
$\overset{N_{O_{n+1}}}{\xleftarrow{\hspace*{2.5cm}}}$&&\\
&&$\overset{N_{O_{n+1}}}{\xrightarrow{\hspace*{2.5cm}}}$\\
$\overset{N_T ,s_i ,m_i}{\xrightarrow{\hspace*{2.5cm}}}$&&\\
&&$\overset{m_i,s_i,N_T,A_v}{\xrightarrow{\hspace*{2.5cm}}}$\\
&&$\overset{ref^{O_n},ref^V}{\xleftarrow{\hspace*{2.5cm}}}$\\
$\overset{m_{i+1},s_{i+1}}{\xleftarrow{\hspace*{2.5cm}}}$&&\\
 \hline
\end{tabular}
\caption{Ownership transfer in ROTIV}
\label{ROTIV}
\end{center}	
\end{figure}

\textbf{Setup:} The issuer $I$ outputs $(q,\mathbb{G}_1,\mathbb{G}_2,\mathbb{G}_T , g_1, g_2, e)$, where $\mathbb{G}_1$, $\mathbb{G}_T$ are subgroups of prime order $q$, $g_1$ and $g_2$ are random generators of $\mathbb{G}_1$ and $\mathbb{G}_2$ respectively, and $e : \mathbb{G}_1 \times \mathbb{G}_2 \rightarrow \mathbb{G}_T$ is a bilinear pairing.\\
The issuer chooses $x \in \mathds{Z}^*_q$ and computes the pair $(g^x_1,g^x_2)$. The $I$'s public and secret keys are:
\begin{equation}\label{issukey}
    sk_I=(x,g^x_1), \ pk_I=g^x_2
\end{equation}
$I$ randomly selects $\alpha_n \in \mathds{Z}^*_q$ and provides each owner $O_n$ with a secret key $sk_{O_n} = \alpha_n$ and a public key $pk_{O_n} =(g^{\alpha^2_n}_1,g^{\alpha_n}_2)$. All owners know each other's public keys.\\\\
\textbf{Tag Initialization: }The issuer $I$ picks a random number $t \in \mathds{F}_q$, where $\mathds{F}_q$ is the finite field with $q$ elements. Using a cryptographic hash function $h :\mathds{F}_q \rightarrow \mathds{G}_1$, $I$ computes $u_0 = 1$ and $v_0 = h^x(t)$. Finally, $I$ chooses randomly a key $k_0 \in \mathds{F}_q$ and stores: $(k_0, s_0)$, where $s_0 = (u_0,v_0)$ into the tag. $I$ also provides $O_n$ with $T$'s information $ref^{O_n}$. This information includes two \emph{dynamic} values $k_{old},k_{new}$ which are updated after each successful transaction and two \emph{static} values $\delta=t,\psi=h^x(t)$ which represent the identification of the issuer of the tag.
\begin{equation}\label{refok}
    ref^{O_n}=(k_{old},k_{new},\delta,\psi)=(k_0,k_{0},t,h^x(t))
\end{equation}
Before accepting the tag, the owner can read the tag and checks the authenticity of the static values of the tag:
\begin{equation}\label{hh}
   e(h(\delta), pk_I) = e(\psi, g_2)
\end{equation}
\textbf{Ownership Transfer: } The ROTIV ownership transfer protocol (Fig.\ref{ROTIV}) is a combination of two mutual authentication sessions between the tag and current and new owners with the ownership transfer protocol between the current owner $O_n$ and the new owner $O_{n+1}$.\\
In $i^{th}$ time instance of the ROTIV protocol:\\
1. New owner $O_{n+1}$ generates a random nonce $N_{O_{n+1}}$ and sends it to the tag and the current owner simultaneously.\\
2. The tag $T$ also generates a random number $N_T$ and send it with its status parameter $s_i = (u_i,v_i)$ and a hash $m_i=h_{k_i}(N_{O_{n+1}},N_T,s_i)$ to the new owner.\\
3. $O_{n+1}$ selects a random number $r_v$ and computes $A_v =u^{r_v}_i$. Then, it sends $N_{O_{n+1}}, N_T, s_i, m_i$ and $A_v$ to the current owner
$O_n$. In this way, $O_n$ is able to authenticate the tag by computing,
\begin{equation}\label{auth1}
    \psi= \frac{v_i}{(u_i)^{\alpha^2_n}}
\end{equation}
Then, it searches in the database to see if $\psi$ is in the database or not. If not, it aborts authentication. Otherwise, it looks up $T$'s ownership references $ref^{O_n}$
in the database to checks if $m_i = h_{k^{new}_i}(N_{O_{n+1}},N_T,s_i)$ or $m_i = h_{k^{old}_i}(N_{O_{n+1}},N_T,s_i)$. For the former case $k_i=k^{new}_i$ and for the latter case $k_i=k^{old}_i$.\\
4. If the authentication process succeeds $O_n$ gives $O_{n+1}$ the following information via a secure channel:\\
\begin{equation}\label{refv}
   ref^V=(A,B,C)=(t,h^x(t),A^{\alpha_n}_v)
\end{equation}
\begin{equation}\label{refo}
    ref^{O_n}=(k_{old},k_{new},\delta,\psi)=(k_i,k_{i+1},t,h^x(t))
\end{equation}
The new owner $O_{n+1}$ check the validity of the provided information by (\ref{hh}).\\
Now, the new owner can verify whether the issuer of the tag $T$ is $I$ by checking whether the following equations hold:\\
\begin{eqnarray}
  e(h(A),pk_I) &=& e(B,g_2) \\
  e(C,g_2) &=& e(A_v,g^{\alpha_n}_2) \\
  e(v_i,g_2)^{r_v} &=& e(B,g_2)^{r_v}e(C,g^{\alpha_n}_2)
\end{eqnarray}
5. If the verification succeeds, $O_{n+1}$ chooses a new random number $r_{i+1}$ and computes:
\begin{equation}\label{cij}
    s_{i+1}=(u_i,v_i)=(g^{r_{i+1}}_1,h^x(t). g^{\alpha^2_nr_{i+1}}_1)
\end{equation}
\begin{equation}\label{mpij}
    m_{i+1}=h_{k_i}(N_T,s_{i+1})
\end{equation}
and sends $s_{i+1},m_{i+1}$ to the tag and updates its database. Now, $T$ authenticates $O_{n+1}$ by checking the content of $m_{i+1}$. If the authentication succeeds $T$ updates its state parameter to $s_{i+1}$ and its symmetric key to the new key $k_{i+1}$ where,
\begin{equation}\label{ki1}
    k_{i+1}=PRNG(k_i,N_{O_{n+1}})
\end{equation}
In order to prevent the current owner from tracing the tag later in the future, the new owner has to run a mutual authentication with the tag outside the range of the current owner after the ownership transfer is complete.
\subsection{Our Attacks}
 In this attack, we target mainly the \emph{ownership privacy} including current and new owner privacy of the ROTIV protocol. Correspondingly, the adversary $\mathcal{A}$ has been one of the owners of the tag $T$ at least once. For example, without loss of generality, we can assume that $\mathcal{A}=O_n$. Therefore, at a time instance e.g. $i$, she has had access to the tags's information $ref^{O_n}$. We also assume that the adversary is passive and thus has access only to \texttt{Execute} and \texttt{Test} queries.\\
 According to the attacking scenario described in Section \ref{scenario}, the adversary follows the procedure below to trace the tag $T$ via distinguishing that which of the two test tags, $T_0$ and $T_1$, are $T$.
\begin{enumerate}
  \item $\mathcal{A}$ retrieves the static information of the tag $T$, $\delta=t,\psi=h^x(t)$, from the information she has been give at time $i$, $ref^{O_n}$.
  \item $\mathcal{A}$ queries \texttt{Test}($j,\mathcal{T}_0,\mathcal{T}_1$) and obtains (\ref{t0}) and (\ref{t1}).
       \begin{equation}\label{t0}
        \{N_{O_l},N_{T_0},m_j, m_{j+1}, s_j, s_{j+1}\}
       \end{equation}
       \begin{equation}\label{t1}
        \{N_{O'_l},N_{T_1},m'_j,m'_{j+1}, s'_j,s'_{j+1}\}
       \end{equation}
which are the messages exchanged between the owner $O_l$ and tags $T_0$ and $T_1$ respectively.\\
\item $\mathcal{A}$ saves $s_j=(u_j,v_j)$ and $s'_i=(u'_j,v'_j)$.
   \item $\mathcal{A}$ checks whether (\ref{1}) or (\ref{2}) holds,
   \begin{eqnarray}\label{ee}
      e(v_j,g_2) &=&e(h(\delta),pk_I)e\left((\frac{v_j}{\psi}),g_2\right)\label{1}\\
       e(v'_j,g_2) &=&e(h(\delta),pk_I)e\left((\frac{v'_j}{\psi}),g_2\right)\label{2}
   \end{eqnarray}
    \item If (\ref{1}) is correct then $\mathcal{A}$ outputs 0 i.e. $T=T_0$, otherwise she outputs 1 i.e. $T=T_1$.
\end{enumerate}
 Note that we can write (\ref{1}) because according to bilinear pairing properties of $e$, we have:
\begin{eqnarray}
  e(v_i,g_2) &=& e(h^x(t).g^{\alpha^2_l r_i},g_2)\nonumber\\
  &=& e(\psi.g^{\alpha^2_l r_i},g_2)\nonumber\\
  &=& e(\psi,g_2)e(g^{\alpha^2_l r_i}_1,g_2)\nonumber\\
   &=& e(h(\delta),g^x_2)e(g^{\alpha^2_l r_i}_1,g_2)\nonumber\\
   &=& e(h(\delta),pk_I)e(g^{\alpha^2_l r_i}_1,g_2)\nonumber\\
   &=& e(h(\delta),pk_I)e\left((\frac{v_i}{\psi}),g_2\right)\nonumber
\end{eqnarray}
Using the scenario above, any owner in the protocol which has had the ownership of the tag $T$ is able to trace it. It is worth mentioning that since the update procedure of state values $s$ are performed independent of their previous values (step 5 of ownership transfer), the aforementioned tracing scenario can be applied both on the state values of the past and the future. Hence any owner who has accessed to the static values of a tag is able to trace it at any time in the past or future by only eavesdropping state parameter of the tag $s$. It implies that the ROTIV protocol lacks both previous owner and new owner privacy properties.\\
\textbf{Remark 1.} It should be noted that if an adversary $\mathcal{A}'$ has access to \texttt{Corrupt} query which gives her this privilege to tamper the tag and access to the tag's static information $t,h^x(t)$, her state of knowledge about the tag is exactly the same as that the adversary $\mathcal{A}$ in the stated attack. Hence, she will also be able to exploit (\ref{ee}) to trace $T$ in any time in the past and future. This implies that the ROTIV protocol lacks forward and backward privacy as well.\\
\section{Chen \emph{et al}'s Protocol}\label{chen2}
Chen \emph{et al.}'s protocol is designed to meet the requirements of EPC Class1 Generation2 standard (ISO18000-6C) for passive RFID tags. According to this standard, RFID tags's computation capabilities is restricted to only performing a 16-bit Cyclic Redundancy Code (CRC) and 16-bit Pseudo-Random Number Generator (PRNG).\\
The authors have claimed that their scheme ensures both forward and backward privacy and it also preserves current and new owner privacy.\\
There are four entities involved in the protocol, a tag $T$, current owner $O_n$, new owner $O_{n+1}$ and issuer $I$ which issues a new issuer identification to be stored into the tags after each ownership transfer phase.
\subsection{Description}
\begin{figure}[b!]
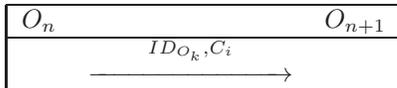

\begin{center}
\begin{tabular}{|l c r|}
\hline
\textbf{$O_n$} &   & \textbf{$O_{n+1}$} \\
\hline
&$\overset{ID_{O_{k}},C_i}{\xrightarrow{\hspace*{2.5cm}}}$&\\
 \hline
\end{tabular}
\caption{Requiring phase}
\label{RF}
\end{center}	
\end{figure}

\begin{figure}[t!]
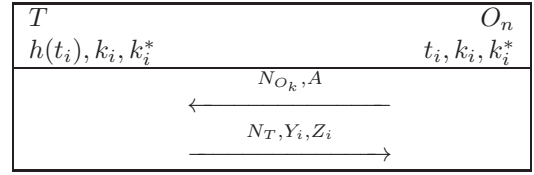

\begin{center}
\begin{tabular}{|l c r|}
\hline
\textbf{$T$} &   &\textbf{$O_n$} \\
$h(t_i),k_i,k^*_i$ &    &$t_i,k_i,k^*_i$ \\
\hline
&$\overset{N_{O_{k}},A}{\xleftarrow{\hspace*{2.5cm}}}$&\\
&$\overset{N_{T},Y_i,Z_i}{\xrightarrow{\hspace*{2.5cm}}}$&\\
 \hline
\end{tabular}
\caption{Authentication phase}
\label{auth}
\end{center}	
\end{figure}

\begin{figure}[t!]
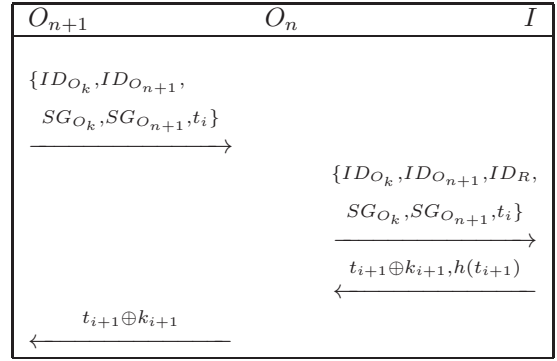

\begin{center}
\begin{tabular}{|l l r|}
\hline
\textbf{$O_{n+1}$} & \textbf{$O_n$}& \textbf{$I$} \\
\hline
&&\\
$\overset{\{ID_{O_{k}},ID_{O_{n+1}},}{}$&&\\
$\overset{SG_{O_{k}},SG_{O_{n+1}},t_i\}}{\xrightarrow{\hspace*{2.5cm}}}$&&\\
&&$\overset{\{ID_{O_{k}},ID_{O_{n+1}},ID_R,}{}$\\
&&$\overset{SG_{O_{k}},SG_{O_{n+1}},t_i\}}{\xrightarrow{\hspace*{2.5cm}}}$\\
&&$\overset{t_{i+1}\oplus k_{i+1}, h(t_{i+1})}{\xleftarrow{\hspace*{2.5cm}}}$\\
$\overset{t_{i+1}\oplus k_{i+1}}{\xleftarrow{\hspace*{2.5cm}}}$&&\\
 \hline
\end{tabular}
\caption{Ownership transfer phase}
\label{otp2}
\end{center}	
\end{figure}

Chen \emph{et al.}'s ownership transfer protocol consist of three phases: \emph{requiring phase}, \emph{authentication phase} and \emph{ownership transfer phase}. In Chen \emph{et al.}'s protocol, the $T$ stores two dynamic symmetric keys $k_i,k^*_i$ and the $h(t_i)$ which is the hash of the issuer identification. In addition to the tag's information the owner has the issuer identification $t_i$.\\
In the $i^{th}$ time instance of requiring phase (Fig.\ref{RF}), the current owner first signs the tag's certificate $t_i$ and the identification of the new owner:\\
\begin{equation}\label{SG}
    SG_{O_n}=Sign_{sk_{Ok}}(t_i,ID_{O_{n+1}})
\end{equation}
After that, it encrypts this message with the next owner's public key to get $C_i$:
\begin{equation}\label{C}
    C_i=E_{pk_{O_{n+1}}}(t_i,SG_{O_n})
\end{equation}
and transfers the message ($ID_{O_{k}},C_i$) to the new owner $O_{n+1}$.\\
In authentication phase (Fig.\ref{auth}), the current owner first generates a random number $N_{O_n}$ and then computes $A_i$:
\begin{equation}\label{A}
    A_i=CRC(k_i\oplus N_{O_n})
\end{equation}
and sends it with $N_{O_n}$ to the tag. Upon receiving these messages, the tag verifies the content of the message $A_i$. If the verification succeeds, the tag generates a new random value $N_T$, and computes the $X_i,Y_i$ and $Z_i$ as following.
\begin{eqnarray}
  X_i &=& CRC(N_T\oplus k^*_i) \\
  Y_i &=& k^*_i\oplus ID_T\oplus X_i\oplus k_{i+1}\\
 Z_i &=& CRC(X_i\oplus k_{i}\oplus Y_i)
 \end{eqnarray}
Moreover, the tag updates its keys as:\\
 \begin{eqnarray}
  k_{i+1} &=& (k^*_i\oplus ID_T\oplus N_T\oplus Y_i) \label{ni1} \\
   k^*_{i+1}&=&PRNG(k^*_{i})
\end{eqnarray}
and transfers ($N_T,Y_i,Z_i$) to the current owner. Upon receiving the message, $O_n$ checks the content of $X_i$ and $Z_i$. If this verification succeeds, it obtains $k_{i+1}$ and updates its values accordingly.\\
In the ownership transfer phase (Fig.\ref{otp2}), the new owner $O_{n+1}$ uses its own private key to decrypt $C_i$ received in the requiring phase and obtains $SG_{O_{k}}$ and $t_i$. Then, it uses the $O_n$'s public key $pk_{O_n}$ to verify the correction of $SG_{O_{k}}$. If the signature is verified successfully, the new owner signs the $ID$ of its own as well as the current owner's:
\begin{equation}\label{ID}
    SG_{O_{n+1}}=Sign_{sk_{Ok+1}}(ID_{O_{k}},ID_{O_{n+1}})
\end{equation}
And sends the tuple $\{ID_{O_{i}},ID_{O_{i+1}}, SG_{O_{i}}, SG_{O_{i+1}}, t_i\}$ to the issuer $I$ to issue a new issuer identification for the tag.\\
The issuer checks the content of this message and if it is correct, it issues the $t_{i+1}$ and computes $t_{i+1}\oplus k_{i+1}$ and $h(t_{i+1})$ and transmits them to $O_n$. Upon receiving this message, $O_n$ sends the former message to the new owner and writes the latter one into the tag's memory. The new owner can also obtain the $t_{i+1}$ by XORing the message received from the current owner and the new key stored in the memory.
\begin{equation}\label{cert}
    t_{i+1}=(t_{i+1}\oplus k_{i+1})\oplus k_{i+1}
\end{equation}
\subsection{Our Attack}
The adversary $\mathcal{A}$ in our attack is one of the previous owners of the tag $T$. Therefore, she has had access to $ID_T, k_i$ and $k^*_{i}$, where the $ID_T$ is the static $ID$ of the tag $T$ or the tag's electronic product code(EPC) and $k_i$ and $k^*_{i}$ are the dynamic keys of the tag at time instance $i$ when the tag has been in the possession of $\mathcal{A}$ as the owner.\\
Being given the messages exchanged between two tags $T_0,T_1$, which one of them is the tag $T$, and another owner $O_l$ at two consecutive time instance $j$ and $j+1$, the adversary follows the procedure below to  distinguish which of the test tags is the tag $T$.
\begin{enumerate}
  \item $\mathcal{A}$ retrieves the static identity of the tag $T$, $ID_T$.
  \item $\mathcal{A}$ queries   \texttt{Test}($j,\mathcal{T}_0,\mathcal{T}_1$),\texttt{Test}($j+1,\mathcal{T}_0,\mathcal{T}_1$)\\
     and obtain\\ $\{A_j,N_{T_0},N_{O_l},Y_j,Z_j\}$,$\{A_{j+1},N'_{T_0},N'_{O_l},Y_{j+1},Z_{j+1}\}$ \\
\begin{eqnarray}
  Y_j &=& k^*_j\oplus ID_{T_0}\oplus X_j\oplus k_j\label{yold} \\
Y_{j+1} &=& k^*_{j+1}\oplus ID_{T_0}\oplus X_{j+1}\oplus k_{j+1}\label{ynew} \\
  Z_j &=& CRC(X_j \oplus k_{j+1}\oplus Y_j)\label{zold}\\
 Z_{j+1} &=& CRC(X_{j+1}\oplus k_j\oplus Y_{j+1})\label{znew}
\end{eqnarray}
From (\ref{yold}), we have:
\begin{equation}\label{ni}
   k_j= k^*_j\oplus Y_j\oplus ID_{T_0}\oplus X_j
\end{equation}
By substituting $k_j$ from (\ref{ni}) in (\ref{znew}), we can write:
 \begin{equation}\label{znew2}
    Z_{j+1} = CRC(k^*_j\oplus ID_{T_0}\oplus X_j \oplus X_{j+1}\oplus Y_j \oplus Y_{j+1})
 \end{equation}
   \item  Now the adversary $\mathcal{A}$ defines the maximum number of iterations as $\tau$ and follows the following steps to determine whether $T_0$ is the tag $T$. It should be noted that the same process can be used to determine whether $T_1$ is the tag $T$.
 \begin{enumerate}
 \item $c=1$
   \item computes:\\
    $k^*=PRNG^c(k^*_i)=\underbrace{PRNG(PRNG(...(k^*_i)..))}_{c\ \mbox{times}}$.
   \item computes\\    $X_j=CRC(k^*\oplus N_{T_0})$ ,\\ $X_{j+1}=CRC(PRNG(k^*)\oplus N'_{T_0})$.
   \item computes $\Delta_X=X_j\oplus X_{j+1}$,$\Delta_Y=Y_j\oplus Y_{j+1}$ .
   \item If $Z_{j+1}\neq CRC(k^*\oplus ID_T\oplus \Delta_X\oplus \Delta_Y)$ and $c< \tau$ then $c=c+1$ and go to b
   \item Else $\mathcal{A}$ outputs 0 i.e. $T_0=T$ and $k^*_{j}=k^*$.
 \end{enumerate}
 \end{enumerate}
This attack shows that the current owner of tag $T$ will be able to trace it at any time in future. Therefore, we can conclude that Chen \emph{et al.}'s protocol lacks new owner privacy.\\
\textbf{ Remark 2.} It should be noted that the procedure above will work when the number of iterations $\tau$ is less than the all possible values for the key $k^*_{j}$. This implies that if the length of key $k^*_{j}$ is $n$, $\tau<<2^n$. So, the tracing process will work efficiently unless the number of passed sessions are comparable to $2^n$.\\
\textbf{ Remark 3.} Any adversary of this kind who has already obtained $k^*_{j}$ from the above procedure is also able to calculate $k_{j+1}$ by (\ref{ni1}).
Then she will be able to extract $t_{i+1}$ from the last message of the tag ownership transfer protocol by using (\ref{cert}). This results in a more dangerous attack in which the current owner is able to even \emph{impersonate} the tag for future interrogations.
\section{Conclusion}\label{conclusions}
In this paper, we investigated the privacy of two ownership transfer protocols. The investigation included the attacks to target the forward and backward privacy as well as previous and new owner privacy properties. Our results showed both protocols are vulnerable to the attacks where the adversary is one of the owners in the system.\\
Any owner in the system as well as any adversary with the capability of tampering the tag are able to trace the tag in the previous and future interrogations in the ROTIV protocol. Therefore, this protocol lacks four stated privacy properties, forward privacy, backward privacy, previous owner privacy and new owner privacy.\\
Chen\emph{ et al.}'s protocol was also shown to be susceptible to the attacks in which the adversary is one of the previous owners of the tag and thus not to fulfil the forward privacy and new owner privacy. This protocol also revealed the whole tag's information to any previous owner and makes the adversary capable of impersonating the tag in further interrogations.

\end{document}